\documentclass[conference]{IEEEtran}
\IEEEoverridecommandlockouts
\usepackage{cite}
\usepackage{amsmath,amssymb,amsfonts}
\usepackage{algorithmic}
\usepackage{graphicx}
\usepackage{textcomp}
\usepackage{xcolor}
\def\BibTeX{{\rm B\kern-.05em{\sc i\kern-.025em b}\kern-.08em
    T\kern-.1667em\lower.7ex\hbox{E}\kern-.125emX}}

\author{\IEEEauthorblockN{1\textsuperscript{st} Ben Dong}
\IEEEauthorblockA{\textit{University of California, Merced} \\
\textit{}
cdong12@ucmerced.edu}
\and
\IEEEauthorblockN{2\textsuperscript{nd} Qian Wang}
\IEEEauthorblockA{\textit{University of California, Merced} \\
\textit{}
qianwang@ucmerced.edu}}

\begin{document}

\title{Evaluating Post-Quantum Cryptography on Embedded Systems: A Performance Analysis\\

}

\vspace{-0.5cm}

\maketitle

\begin{abstract}
The National Institute of Standards and Technology (NIST) has finalized the selection of post-quantum cryptographic (PQC) algorithms for use in the era of quantum computing. Despite their integration into TLS protocol for key establishment and signature generation, there is limited study on profiling these newly standardized algorithms in resource-constrained communication systems. In this work, we integrate PQC into both TLS servers and clients built upon embedded systems. Additionally, we compare the performance overhead of PQC pairs to currently used non-PQC schemes.
\end{abstract}

\vspace{-0.3cm}
\section{Introduction}
With the computational superiority of the quantum computer, all currently deployed cryptographic algorithms based on the integer factorization problem (e.g., RSA), the discrete logarithm problem (e.g., DH), or the elliptic-curve discrete logarithm problem (e.g., ECC) will be broken in polynomial time. As a result, it is imperative to update/upgrade the current crypto algorithms to those Post-quantum crypto (PQC) schemes. The National Institute of Standards and Technology (NIST) recently announced the finalists for key exchange and digital signature schemes for standardization \cite{moody2021nist}. Among the four selected schemes, CRYSTALS-kyber \cite{bos2018crystals} is selected to exchange keys among systems, and CRYSTALS-dilithium \cite{ducas2018crystals}, falcon\cite{fouque2018falcon}, and SPHINCS+ \cite{bernstein2019sphincs+} are the three signature schemes selected to process the signature generation and verification respectively.
Given the inherent backend differences among these algorithms, the logical next step involves evaluating and comparing the performance and overhead of various standardized PQC algorithms. Furthermore, in security applications involving resource-constrained devices such as IoT devices, the timely transmission of encrypted data is paramount. Consequently, assessing the performance of PQC algorithms on embedded, low-power devices becomes crucial for advancing state-of-the-art applications in this domain.

TLS has emerged as a widely utilized protocol for securely transmitting encrypted messages among various parties. Post-quantum key establishments and signature-generating schemes also have been successfully integrated into TLS protocol to assist the migration of the PQC era \cite{sikeridis2020post}. However, there is less focus on the embedded systems, especially when the embedded system itself serves as the server or host. In this work, we aim to establish the PQC-TLS on a distributed embedded system and profile the performance among different PQC algorithms when the device serves as both clients/hosts. We are focusing on saturating the computation overhead on the server’s side rather than on the client side. That differences our work with prior arts \cite{burstinghaus2020post} as we have multiple clients to communicate with the server (on an embedded system) simultaneously.  

\vspace{-0.3cm}
\section{Methodology}

\subsection{Overview of Post-quantum Crypto}
Post-quantum cryptography (PQC) refers to those cryptographic algorithms designed to be resistant against attacks facilitated by the quantum computer. PQC encompasses various cryptographic families, including lattice-based, hash-based, code-based, multivariate, and isogeny schemes. Notably, hash-based and lattice-based schemes have been selected by standardization organizations such as NIST, considering factors such as performance, security, and applicability. Among those diverse PQC algorithms, hash-based schemes are more mature than others as they have undergone extensive study decades ago and are well understood in terms of their security properties. For instance, before NIST chose SPHINCS+ for standardization, the other similar hash-based schemes like XMSS and LMS had already found applications in tasks such as firmware updates and security signing. Lattice-based cryptography delves into the study of algebraic structures such as Ring Learning With Errors (Ring-LWE) or Short Integer Solution (SIS) of latices. These approaches are particularly advantageous in terms of efficiency, especially in implementing key-exchange frameworks, as they entail relatively lower computational overhead. Indeed, NIST's standardization process has acknowledged the efficiency of lattice-based cryptography, choosing three key schemes: Kyber for key encapsulation, and Dilithium and Falcon for signature generation. This selection highlights the strength and suitability of lattice-based approaches for resource-constrained systems in the post-quantum era.



\subsection{Applying PQC on TLS protocol}


Typically, TLS employs ECDH, an elliptic curve variant of Diffie-Hellman, for key exchange. However, adjustments to both the protocol and contents are required to incorporate the Key exchange scheme using PQC into the TLS protocol. The TLS handshake begins with the client sending a \texttt{ClientHello} message, specifying supported cipher suites that define cryptographic algorithms for various purposes. In the PQC scenario, these cipher suites need to include the Kyber for key exchange, as well as Dilithium, Falcon, and SPHINCS+ for signature generation. Upon receiving the \texttt{ClientHello} message, the server responds with a \texttt{ServerHello} message, confirming the chosen cipher suite for key exchange and signature generation. The server also provides its certificate and any necessary parameters for the chosen PQC schemes (e.g., public keys for signature verification). The key exchange process (i.e., the encapsulation on the random to produce a shared secret) is then performed using the Kyber scheme, generating a shared secret key between the client and the server.  After the key exchange, signatures are generated using the chosen PQC schemes (Dilithium, Falcon, or SPHINCS+) to authenticate the server's identity. The data fields, including the certificate, shared secret, and random values, are signed using PQC schemes for authentication and integrity assurance. This package, comprising the certificate and the shared secret with its signature, is transmitted back to the client. At the client's end, the received signature is first verified to ensure its authenticity and integrity. Subsequently, the client performs a decapsulation operation to extract the shared secret from the ciphertext. With both the client and server possessing the shared secret, they can independently derive the necessary session keys for securing subsequent communication. By incorporating these adjustments, the TLS handshake protocol can effectively utilize post-quantum cryptographic schemes for key exchange and signature generation.

\section{Experimental Results}

\subsection{Experimental Set-up}
\vspace{-0.15cm}
Our experiments were conducted on a Raspberry Pi system, featuring an ARM Cortex core, with a maximum operating frequency of 2.4 GHz. In our setup, the Raspberry Pi system serves as both the server and client to assess its capability in handling PQC computations within the TLS framework. Unlike conventional setups where embedded systems typically serve as clients only, our approach allows us to comprehensively profile all computations involved in the TLS handshake process. This includes functions such as encapsulation (\texttt{encap()}) and signature generation (\texttt{sig()}) on the server side, as well as decapsulation (\texttt{decap()}) and signature verification (\texttt{verify()}) on the client side. By utilizing the Raspberry Pi system in this dual role, we can effectively evaluate the performance and efficiency of post-quantum cryptographic operations within the TLS handshake, taking into account the computational constraints and resource limitations inherent to embedded systems. In our implementation, we utilize OpenSSL, a widely-used TLS library framework, and integrate the PQC algorithms from the specialized quantum cryptography library as the liboqs. This combination allows us to seamlessly incorporate PQC schemes into the TLS handshake protocol. 



\subsection{Results and Analysis }
In our evaluation, we measured the performance of the TLS handshake by determining the number of connections per second that could be established between the server and the clients. For our control group, we utilized the default ECDHE-x25519 for key exchange and RSA for signature generation. To assess the impact of PQC schemes on handshake performance, we calculated the ratio of connections per second achieved with each PQC pair to those achieved with the control group which provides insights into the relative efficiency of PQC schemes.

Based on our evaluation results, Kyber-Falcon pairs exhibit the best performance among the PQC schemes, achieving the highest ratio compared to the control group. The efficiency is attributed to Falcon's smaller signature size, resulting in less overhead in communication time. Kyber-SPHINCS+ pairs demonstrate the lowest performance, with less than 30\% of the control group's efficiency, due to SPHINCS+'s extensive computation on numerous underlying hash function calls. Additionally, SPHINCS+ has the longest signature size, requiring more time for transmission over the TLS handshake. Kyber-Dilithium pair shows moderate performance, with the potential for further improvement through associated cryptographic acceleration, particularly for operations such as Number Theoretic Transform (NTT). Notably, Dilithium's performance improvement prospects are more promising compared to Falcon, as the latter requires floating-point calculations for acceleration, which may present additional challenges. These findings highlight the trade-offs between performance and cryptographic strength associated with different PQC schemes. While Falcon excels in terms of performance due to its smaller signature size, Dilithium shows potential for enhancement through cryptographic acceleration. Meanwhile, SPHINCS+ demonstrates the lowest performance due to its computational overhead and longer signature size.
\vspace{-0.3cm}
\begin{figure}[!htbp]
\centerline{\includegraphics[width=1\columnwidth]{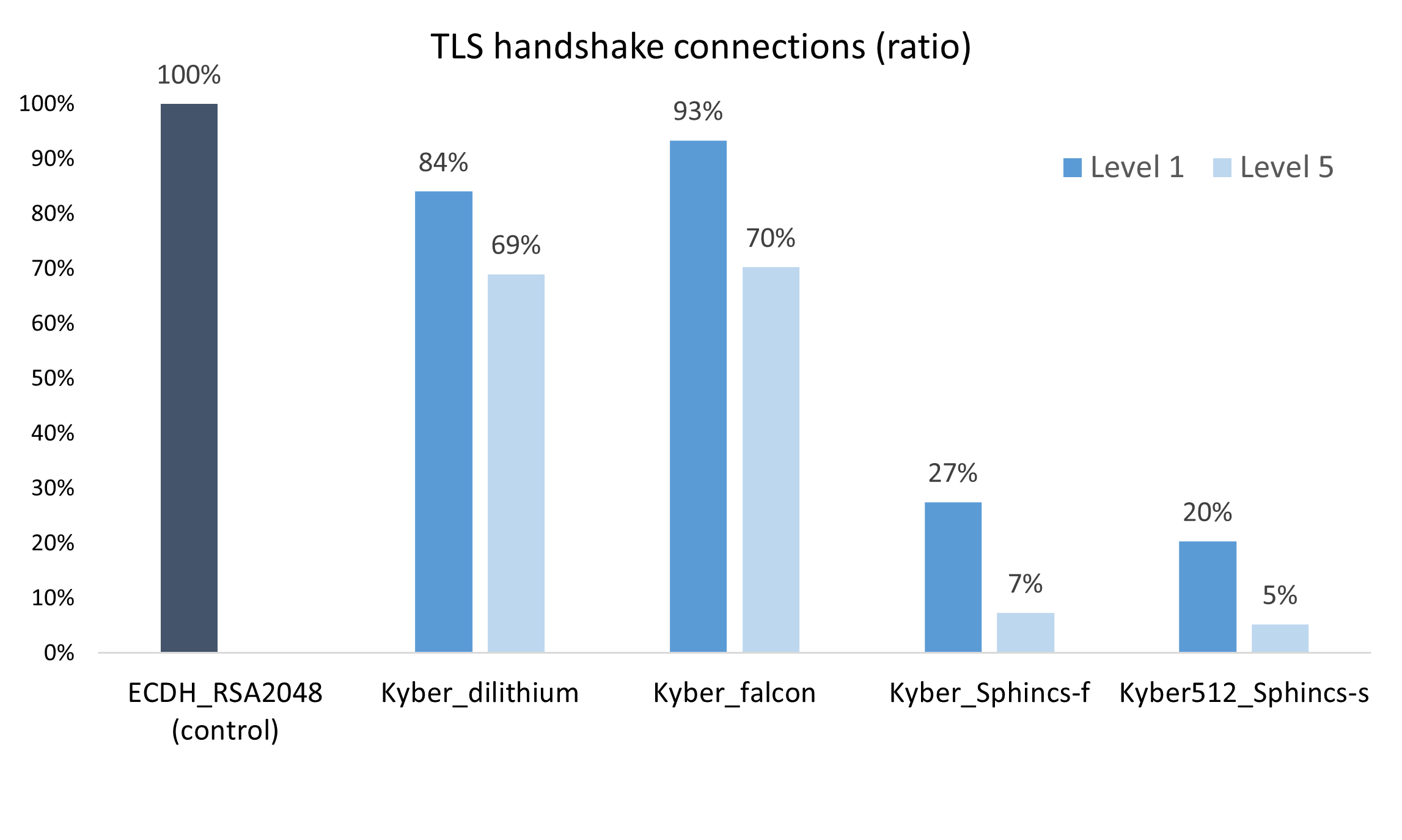}}
\vspace{-0.6cm}
\caption{TLS handshake connections over PQC KEM-SIG pairs}
\label{fig}
\end{figure}
\vspace{-0.4cm}
\section{Conclusion}
The primary contribution of this paper lies in the thorough profiling of PQC-TLS performance on resource-constrained devices. We conducted experiments using a Raspberry Pi system, serving both as the server and client in the TLS handshake process. Our results provide insights into the strengths and weaknesses of various PQC schemes. Moving forward, we aim to explore accelerated variants of PQC algorithms tailored for embedded systems.

\bibliographystyle{IEEEtran.bst}
\bibliography{reference}
\end{document}